\documentclass[conference,9pt,dvipsnames]{IEEEtran}

\AtBeginDocument{%
  \providecommand\BibTeX{{%
    \normalfont B\kern-0.5em{\scshape i\kern-0.25em b}\kern-0.8em\TeX}}}
\usepackage{cite}

\usepackage{url}
\usepackage{enumitem}

\usepackage[utf8]{inputenc} 
\usepackage[T1]{fontenc}    
\usepackage{url}            
\usepackage{booktabs}       
\usepackage{amsfonts}       
\usepackage{amsmath}
\usepackage{nicefrac}       
\usepackage{microtype}      
\usepackage{xcolor}         
\usepackage{colortbl}
\usepackage{graphicx}
\usepackage{svg}
\usepackage{multirow}
\usepackage{wrapfig}
\usepackage{threeparttable}
\usepackage{tikz}
\usepackage{makecell}
\usepackage{tablefootnote}
\usepackage[hidelinks]{hyperref}

\usepackage{color}
\usepackage{xspace}
\usepackage{caption}
\usepackage{adjustbox}

\usepackage{subfigure}

\usepackage{amsmath}
\usepackage{amssymb}
\usepackage{mathtools}
\usepackage{amsthm}
\usepackage{hyperref}
\usepackage{adjustbox}
\usepackage{xcolor}
\theoremstyle{plain}

\theoremstyle{definition}

\theoremstyle{remark}

\begin{document}
\pagestyle{plain}
\title{\name: When Agentic Context Evolution Meets RTL-Specialized LLMs}



\author{
\IEEEauthorblockN{Chenhui Deng, Zhongzhi Yu, Guan-Ting Liu, Nathaniel Pinckney, 
Brucek Khailany, Haoxing Ren$^{*}$}
\IEEEauthorblockA{
NVIDIA \\
\{cdeng, zhongzhiy, dannliu, npinckney, bkhailany\}@nvidia.com, markren@agentrys.ai
}
}


\newcommand{\name}{ACE-RTL\xspace}

\newcommand{\red}[1]{\textcolor{red}{#1}}
\newcommand{\cd}[1]{\textcolor{violet}{[chenhui: #1]}}


\maketitle
\begingroup
\renewcommand{\thefootnote}{\fnsymbol{footnote}}
\footnotetext[1]{Work done during affiliation with NVIDIA, now at Agentrys.}
\endgroup

\begin{abstract}
Recent advances in LLMs have sparked growing interest in applying them to hardware design automation, particularly for accurate RTL code generation. Prior efforts follow two largely independent paths: \textit{(i)} training domain-adapted RTL models to internalize hardware semantics, \textit{(ii)} developing agentic systems that leverage frontier generic LLMs guided by simulation feedback. However, these two paths exhibit complementary strengths and weaknesses. 
In this work, we present ACE-RTL that unifies both directions through Agentic Context Evolution (ACE). ACE-RTL integrates an RTL-specialized LLM, trained on a large-scale dataset of 1.7 million RTL samples, with a frontier reasoning LLM through three synergistic components: the generator, reflector, and coordinator. These components iteratively refine RTL code toward functional correctness. 
We further analyze a parallel scaling strategy that reduces wall-clock
iterations to first success by exploring diverse debugging trajectories
concurrently.
On the CVDP benchmark, ACE-RTL achieves up to a $41.02\%$ pass rate improvement over $14$ competitive baselines.
\end{abstract}

\section{Introduction}

Large language models (LLMs) have shown rapid progress in software development and are increasingly explored for hardware design automation~\cite{jimenez2023swe, cursor2025, claudecode2025, liu2023chipnemo, deng2025chipalign}. As hardware systems continue to grow in scale and complexity, developing register-transfer-level (RTL) designs remains a labor-intensive and error-prone process requiring deep domain expertise. Leveraging LLMs to translate natural-language intent into functionally correct RTL code can significantly accelerate design cycles. Moreover, accurate RTL generation is essential for downstream workflows such as performance optimization and architectural design space exploration~\cite{zhang2025aspen, yao2024rtlrewriter, wang2025symrtlo}. 

\begin{figure}[t!]
\begin{center}
	\includegraphics[width=\columnwidth]{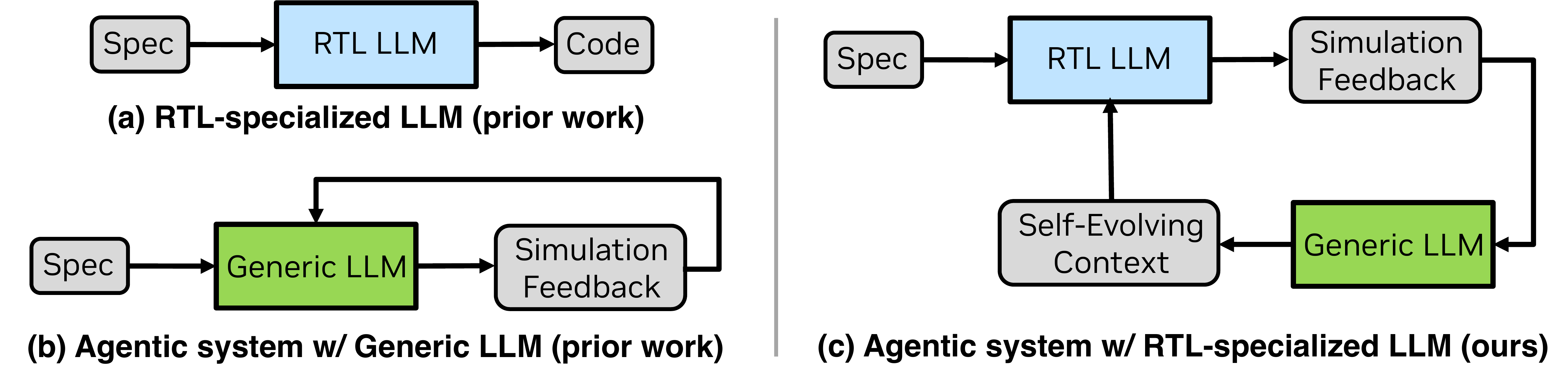}
\vspace{-18pt}
	\caption{Comparison of prior works and \name.} 
\vspace{-21pt}
     \label{figure:motif}
\end{center}
\end{figure}

In parallel to this motivation, the research community has pursued two main directions, as shown in Figures \ref{figure:motif}(a) and \ref{figure:motif}(b), for applying LLMs to RTL code generation. \textbf{\textit{(i)}} The first focuses on developing domain-specialized LLMs trained on RTL data, where models are fine-tuned to internalize RTL syntax and hardware design semantics. For instance, RTLCoder~\cite{liu2024rtlcodera, liu2024rtlcoderb} develops a specialized model fine-tuned on a dedicated Verilog dataset. CraftRTL~\cite{liu2024craftrtlhigh} further enhances RTL coding capability by training on synthetic data that capture non-textual design representations. More recently, ScaleRTL~\cite{deng2025scalertl} scales up post-training on a large RTL reasoning corpus, yielding a strong specialized model that achieves state-of-the-art (SoTA) results on mainstream RTL benchmarks. \textbf{\textit{(ii)}} The second direction centers on agentic systems that leverage frontier generic LLMs together with RTL-specific tools such as simulators and waveform analyzers. Along this line, VerilogCoder~\cite{Ho2025verilogcoder} introduces a multi-agent framework based on GPT4-Turbo~\cite{openai2024gpt4turbo} that decomposes RTL design into fine-grained sub-tasks through a task-and-circuit relation graph and repairs functional errors using a waveform tracing tool. Similarly, MAGE~\cite{zhao2024mage} employs multiple Claude~\cite{anthropic2024claude} LLM agents tightly integrated with hardware-domain toolchains to produce functionally correct RTL code.

Despite these advances, the two research directions remain largely orthogonal and exhibit complementary limitations. Training-based approaches often yield RTL-specialized models with limited general capabilities such as long-context reasoning, multi-step planning, and instruction following, reducing their efficacy on complex design tasks. Conversely, agent-based methods lack hardware knowledge learned from large-scale RTL datasets, causing the system to struggle on problems that require deep hardware semantics and domain expertise. These limitations highlight the need for a unified framework that integrates an RTL-specialized model with a frontier general-purpose LLM, enabling the system to leverage the strengths of both paradigms.

In this work, we advance LLM-based RTL coding by synergizing the complementary strengths of an RTL-specialized model and a frontier general-purpose LLM. To this end, we propose \name, which adopts the notion of Agentic Context Evolution (ACE) through three components: the Generator, the Reflector, and the Coordinator. As illustrated in Figure~\ref{fig:overview-parallel}(a), the Generator is an RTL-specialized LLM trained on 1.7 million diverse RTL samples. 
Different from prior RTL models that are trained exclusively on specification-to-RTL data, our dataset additionally includes code editing and debugging tasks. 
This design better prepares the Generator for its role in the agentic loop, where it not only generates initial RTL code from specification, but also edits and fixes bugs in code in later iterations. 
When errors occur, the Reflector, powered by the frontier LLM Claude4-Sonnet \cite{anthropic2025claude4}, analyzes simulation feedback, traces errors to faulty logic, and provides high-level fix guidance.
The Coordinator then incorporates this guidance into an incrementally evolving context that guides the Generator in the next iteration of code generation. 
By iteratively cycling among the Generator, Reflector, and Coordinator, \name effectively combines domain-specific RTL expertise with strong general-purpose reasoning to produce functionally correct RTL designs.
We further introduce a parallel scaling strategy that launches multiple \name processes simultaneously and terminates the remaining ones once any process succeeds. This largely reduces the iterations required to reach a correct RTL implementation, with only a modest increase in token usage.

To evaluate the efficacy of \name, we compare it against the strongest frontier LLMs, as well as SoTA RTL-specialized models and agents. More importantly, instead of relying on prior RTL benchmarks such as VerilogEval~\cite{liu2023verilogeval, pinckney2024verilogeval} and RTLLM~\cite{lu2024rtllm}, which primarily consist of relatively simple specification-to-RTL generation tasks (e.g., small modules or textbook-style problems), we conduct evaluations on the Comprehensive Verilog Design Problems (CVDP) benchmark~\cite{pinckney2025cvdp}. CVDP encompasses hundreds of diverse and challenging RTL coding tasks.
Since earlier benchmarks have become saturated and no longer sufficiently differentiate recent approaches, we argue that evaluating on CVDP provides a more rigorous and practical assessment of model capability and more convincingly demonstrates meaningful progress in this domain. We summarize our technical contributions as follows:

\noindent $\bullet$ To our knowledge, \name is the first approach to integrate an RTL-specialized model into an agentic system through iterative context evolution, demonstrating that a finetuned RTL model can contribute not only as a strong standalone generator, but also as an effective core component in iterative agentic refinement.

\noindent $\bullet$ 
Trained on a 1.7M-sample RTL corpus spanning diverse tasks, our RTL model achieves up to a $21.45\%$ higher pass rate than strong standalone baselines such as Claude4 and GPT5, while also aligning well with the iterative bug-fixing setting used in our agentic loop.

\noindent $\bullet$ By combining this RTL model with the ACE framework, \name achieves up to a $41.02\%$ pass-rate improvement over all baselines on the CVDP benchmark, demonstrating the effectiveness of iterative context evolution for complex RTL tasks.

\noindent $\bullet$ With parallel scaling, \name speeds up convergence to correct RTL code, achieving a $2.77\times$ reduction in wall-clock iterations relative to the non-scaled version, with only a modest increase in token usage.
\section{Background}
\subsection{Domain-Adapted LLMs for RTL Tasks}
\label{llm}
Adapting LLMs for RTL coding via model training has attracted growing attention due to their potential to improve hardware design productivity. 
RTLCoder~\cite{liu2024rtlcodera, liu2024rtlcoderb} develops a GPT-assisted pipeline for automated RTL data synthesis, while CraftRTL~\cite{liu2024craftrtlhigh} generates synthetic, correct-by-construction samples derived from design artifacts. OriGen~\cite{cui2024origen} adopts compiler-guided reflection and augmentation to improve functional correctness. Recently, 
ScaleRTL~\cite{deng2025scalertl} scales both training data and test-time compute by fine-tuning on billions of RTL reasoning tokens and applying iterative reasoning extension, achieving SoTA results on VerilogEval~\cite{pinckney2024verilogeval} and RTLLM~\cite{lu2024rtllm} benchmarks. These efforts collectively advance the development of RTL-specialized models. 
However, those prior models are trained predominantly on specification-to-RTL generation data, with limited coverage of code editing and debugging behaviors. This training mismatch makes them less effective in practical agentic settings, where the model needs to iteratively repair buggy RTL produced in earlier iterations. 
To tackle this issue, we curate a 1.7M-sample RTL corpus covering code generation, editing, and debugging tasks. Training on this corpus yields an RTL-specialized model that is not only strong at standalone RTL generation, but also fits agentic workflows for iterative bug fixing.

\subsection{Agentic Systems and Context Evolution}
\label{agent}
Recent agentic systems such as VerilogCoder~\cite{Ho2025verilogcoder} and MAGE~\cite{zhao2024mage} have demonstrated the potential of multi-agent collaboration for RTL code generation. VerilogCoder decomposes RTL design into fine-grained subtasks using a Task and Circuit Relation Graph (TCRG) and employs a waveform-tracing debugger to locate functional errors, while MAGE leverages multiple specialized agents for RTL generation, testbench creation, and debugging through simulator-based feedback. Despite their promising results, both methods depend heavily on complex, tool-specific planning and lack an RTL-specialized model that captures broad hardware coding logic, which can make them harder to generalize beyond predefined task structures. 
In this work, we take a different perspective: we argue that the key role of an agentic system is to dynamically construct the right context to guide the RTL generator toward correct solutions. To this end, \name iteratively evolves contextual information from previous iterations to progressively steer an RTL model toward functional correctness.

\subsection{RTL Benchmarks for LLM Evaluation}
\label{bench}
Prior mainstream benchmarks such as VerilogEval and RTLLM mostly evaluate small, self-contained RTL problems, emphasizing short specification-to-RTL snippets rather than sustained reasoning over realistic design contexts. RTL-Repo~\cite{allam2024rtlrepo} takes a step toward practical settings by introducing GitHub-derived projects and prompting models to complete redacted regions, yet it remains limited to code completion and omits broader challenges such as specification-to-RTL generation, code modification, or debugging. We argue that rigorous assessment requires benchmarks stressing both context length and task diversity. Therefore, this work focuses on evaluating RTL methods on CVDP~\cite{pinckney2025cvdp}, whose problem statements and target code are several orders of magnitude longer than those in VerilogEval. CVDP includes hundreds of complex RTL problems spanning multiple tasks and hardware domains such as arithmetic units, on-chip protocols, memory hierarchies, DSP kernels, and communication blocks. Empirically, we observe that CVDP better distinguishes model capability and more faithfully reflects the demands of practical RTL design.
 
\section{The \name Framework}
\label{method}
\begin{figure*}[t!]
\begin{center}
	\includegraphics[width=0.94\textwidth]{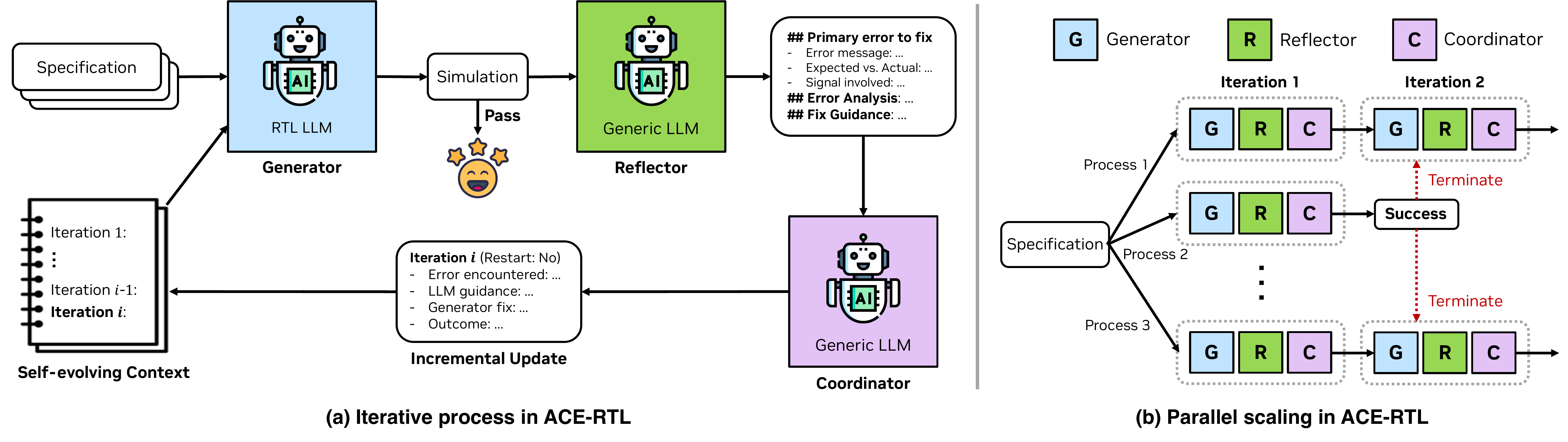}
    \vspace{-6pt}
	\caption{Overview of \name and its parallel scaling strategy.} 
 \protect\label{fig:overview-parallel}
 \vspace{-20pt}
\end{center}
\end{figure*}
Figure \ref{fig:overview-parallel} provides an overview of ACE-RTL, which integrates an RTL-specialized model and a frontier generic LLM through three synergistic components: the Generator, Reflector, and Coordinator. At a high level, ACE-RTL operates in an iterative agentic loop that progressively refines RTL code toward functional correctness. 
These three components are discussed in detail in Sections \ref{sec:generator}, \ref{sec:reflector}, and \ref{sec:curator}, respectively.
In addition, Section~\ref{sec:parallel} introduces a parallel scaling strategy that executes multiple \name instances concurrently to speedup convergence efficiency.

\subsection{Generator: An RTL-Specialized LLM}
\label{sec:generator}
We develop an RTL-specialized model, referred to as \name-Generator, that is trained on a large-scale synthetic dataset to capture diverse hardware design patterns and semantics. 
Unlike prior synthetic data generation (SDG) approaches that focus primarily on specification-to-RTL translation, our SDG considers a broader family of tasks that better reflect real-world hardware development workflows, including \emph{code editing} and \emph{debugging}. These tasks enable the model to learn refinement and error-correction capabilities that are essential for RTL design, particularly when deployed within an agentic workflow.

Specifically, we construct three categories of training samples under a unified formulation that transforms an input condition into a correct (golden) RTL implementation $C^\star$. The input condition may consist of a specification alone or a specification paired with an existing design that must be refined or corrected. We denote the input as $X$ and define the task as:
$X \rightarrow C^\star$,
where $X$ takes different forms:
\begin{equation}
X =
\begin{cases}
S & \text{(Spec-to-RTL)} \cr
(S, C_{\text{base}}) & \text{(Editing)} \cr
(S, C_{\text{bug}}) & \text{(Debugging)}
\end{cases}
\end{equation}

Here, $S$ denotes the specification, which includes functional requirements and, depending on the task, may additionally include feature requests or diagnostic information such as signal mismatches. $C_{\text{base}}$ represents a partial implementation missing required functionality, while $C_{\text{bug}}$ denotes a faulty implementation containing design errors. Both editing and debugging tasks therefore require the model to reason over existing code context and perform targeted transformations, rather than synthesizing designs from scratch. In the following, we describe the steps of our SDG pipeline in detail.


\paragraph{\textbf{High-Quality RTL Seed Collection}}
We begin by collecting approximately five million RTL scripts from public repositories and open hardware projects. To obtain a clean pool of seed designs, we apply deduplication, remove machine-generated artifacts (e.g., netlists), filter extreme outliers (30--2{,}000 lines per module), and validate syntax using \textit{Icarus Verilog (iverilog)}~\cite{williams2002icarus}. We further perform benchmark decontamination using the Jaccard similarity metric~\cite{deng2025scalertl}, discarding any sample with similarity greater than $0.8$ to evaluation benchmarks. After these stages, we retain 157K high-quality RTL scripts, which we use as \emph{seed implementations} to prompt LLMs for synthesizing diverse RTL problems and corresponding golden solutions. 

\paragraph{\textbf{Specification-to-RTL Data Construction}}
Unlike prior works that rely on chain-of-thought (CoT) reasoning supervision, we intentionally avoid training on CoT traces, since they substantially increase inference latency when the model is deployed within an agentic loop. Instead, we use a small set of human-crafted specification-to-RTL examples only as in-context demonstrations to teach frontier LLMs the desired task format and complexity. The large-scale synthetic corpus is then constructed from the collected RTL seeds in Step (a): for each seed RTL implementation, we prompt frontier LLMs, including GPT-OSS-120B and DeepSeek-R1, to generate diverse synthetic specification–code pairs conditioned on the seed design.

\paragraph{\textbf{Editing and Debugging Data Construction}}
Beyond specification-to-RTL generation, we further construct editing and debugging samples to model realistic refinement workflows. Starting from the collected RTL seed implementations, which are treated as golden designs for editing and debugging tasks, LLMs are prompted to generate intermediate variants, including simplified versions with missing functionality and buggy versions with injected design errors. Simplified variants are created by removing advanced features (e.g., pipeline stages), replacing optimized control logic with baseline implementations, or restricting corner-case handling such as overflow detection or boundary conditions. Buggy variants are produced by introducing realistic RTL faults, including incorrect state transitions in finite-state machines, timing or handshake protocol violations, off-by-one counter logic, and related design mistakes.

Given these intermediate implementations, LLMs then generate corresponding specifications that either describe the required feature extensions (for editing) or provide diagnostic information (for debugging). Diagnostic specifications are constructed by analyzing the injected faults in the buggy implementation and generating failure descriptions consistent with the underlying errors. The resulting training instances require recovering the correct implementation from incomplete or faulty inputs, encouraging the model to learn iterative refinement and error-correction behaviors.

\paragraph{\textbf{Filtering and Quality Control}}
All generated samples undergo multi-stage validation to ensure correctness and diversity. 
We first perform syntax validation and benchmark decontamination to ensure structural correctness and prevent data leakage~\cite{deng2025scalertl}~\cite{yang2023rethinking}. We then apply semantic alignment scoring using an LLM-as-Judge framework~\cite{gu2024survey}, which evaluates the consistency between specifications and implementations on a 1–5 scale. To construct the evaluation framework, we first apply the judge to a subset of training samples and manually verify whether the assigned scores are reasonable. If inconsistencies are observed, we refine the evaluation prompt and repeat the process until the scoring behavior is satisfactory. The finalized judge is then applied to the full dataset, and samples with scores below 3 are 
discarded to maintain data quality.
This strategy provides an efficient way for large-scale SDG while maintaining sufficient quality control. Through this pipeline, we curate approximately 1.7 million high-quality training instances spanning diverse RTL coding problems.

Using this large-scale RTL dataset, we apply supervised fine-tuning (SFT)~\cite{chung2024scaling} to obtain the RTL-specialized model. The resulting model captures rich hardware domain knowledge and RTL semantics, enabling accurate code generation across diverse design tasks. Moreover, training on editing and debugging supervision enables the model to effectively analyze, refine, and correct RTL implementations, making it well suited for deployment within our agentic framework.

\subsection{Reflector: A Frontier Reasoning LLM}
\label{sec:reflector}
While our Generator produces and refines implementations within the agentic loop, effective iteration also requires interpreting simulation outputs and identifying corrective actions. We therefore introduce the Reflector, a generic frontier LLM (Claude4-Sonnet) that analyzes simulation feedback and long-context error reports to guide subsequent refinement. Processing such complex information benefits from the broader reasoning capability and context capacity of frontier models.

At each iteration, we invoke \textit{iverilog} once to compile and execute the
candidate RTL against the benchmark test harness. When the run fails, the
resulting logs are parsed into a structured report capturing the error message
together with the expected-versus-actual signal behavior. Claude4-Sonnet then
serves as the Reflector’s reasoning engine: it combines this structured feedback
with the design specification and the current RTL implementation to produce a
concise diagnosis and high-level fix guidance. This output is then
passed to the Coordinator to construct the evolving context for subsequent
refinement steps.
Notably, the entire process depends only on \textit{iverilog}, a lightweight and open-source simulator, rather than complex proprietary toolchains. 

\subsection{Coordinator: A Context Evolution Engine}
\label{sec:curator}
The Coordinator in \name serves as the memory and control center of the agentic loop, maintaining a dynamic, self-evolving context that guides the Generator in subsequent iterations. Its primary role is to provide an incrementally refined context that captures an organized record of prior attempts, enabling more informed reasoning while avoiding redundant or counterproductive fixes. This component is implemented through efficient scripts, with Claude4-Sonnet providing core reasoning support, to continuously update and restructure the evolving context based on the system’s progress.

The Coordinator performs two key functions. First, it aggregates debugging history across iterations, tracking identified errors, suggested fixes, and their outcomes. This history is incorporated into the evolving context, preserving useful progress and preventing regressions so that the Generator can focus on unresolved failures.

Second, the Coordinator monitors progress across iterations. It uses Claude4-Sonnet to examine the debugging trajectory and determine whether the current search has stalled. When the same primary failure persists across several iterations without meaningful improvement, the Coordinator triggers a \textit{restart mechanism}, discarding the current implementation and prompting the Generator to produce a fresh RTL design. This leverages LLM stochasticity to explore alternative implementations that are easier to correct.

By combining cross-iteration history aggregation with an adaptive
restart mechanism, the Coordinator keeps the evolving context informative and flexible. It preserves useful debugging insights when progress is made and triggers controlled resets during stagnation,
enabling efficient convergence to correct RTL code.

Notably, from the Generator’s perspective, the self-evolving context resembles an augmented specification that describes failure symptoms or required modifications. Because the Generator is trained on both editing and debugging tasks, it is naturally capable of interpreting such context, without requiring additional model adaptation. 

\subsection{Parallel Scaling in \name}
\label{sec:parallel}
Although the iterative debugging framework in \name effectively improves code correctness, it may require a large number of iterations to converge when starting from an unfavorable initial RTL generation. To address this inefficiency, we introduce a \textit{parallel scaling} strategy that exploits the inherent stochasticity of LLM-based code generation to accelerate convergence.
Instead of relying on a single debugging trajectory, the parallel scaling strategy launches multiple \name processes simultaneously, as shown in Figure \ref{fig:overview-parallel}(b). Each process begins with a distinct RTL implementation generated by the Generator from the same input specification. These processes then proceed separately through the iterative loop of the Generator, Reflector, and Coordinator components, performing debugging and refinement in parallel.

During execution, all processes are monitored concurrently. Once any process produces an RTL implementation that passes all functional test cases, the system immediately terminates all remaining processes and outputs the correct code. If none of the processes achieve functional correctness within the predefined maximum number of iterations, all are stopped automatically.
This strategy leverages LLM generation diversity to reduce time-to-first-success with a trade-off in total inference-time compute.
In Section~\ref{para_scale}, we evaluate parallel scaling from both convergence and inference-cost (i.e., token usage) perspectives. Our goal is not to claim lower total compute, but to show that parallel scaling provides a favorable speed-cost trade-off. 
\section{Experiment}

\subsection{Experimental Setup}
\label{setup}
\begin{table*}[t!]
\centering
\caption{Functional correctness on CVDP. We report \textbf{Pass@1} and \textbf{Agentic Pass Rate (APR)}. For agentic methods, Pass@1 is omitted (“---”) since it equals APR. \textbf{Bold} indicates our best results, and \underline{underline} the strongest baseline in each category.}
\vspace{-5pt}
\label{tab:cvdp}
\renewcommand{\arraystretch}{1.0}
\setlength{\tabcolsep}{5.5pt}
\begin{tabular}{llcccccccc}
\toprule
\multirow{2}{*}{\textbf{Type}} & \multirow{2}{*}{\textbf{Model}} &
\multicolumn{2}{c}{\makecell{\textbf{CVDP-cid002}\\ \footnotesize Code Completion}} &
\multicolumn{2}{c}{\makecell{\textbf{CVDP-cid003}\\ \footnotesize Spec-to-RTL}} &
\multicolumn{2}{c}{\makecell{\textbf{CVDP-cid004}\\ \footnotesize Code Modification}} &
\multicolumn{2}{c}{\makecell{\textbf{CVDP-cid016}\\ \footnotesize Code Debugging}} \\
\cmidrule(lr){3-4} \cmidrule(lr){5-6} \cmidrule(lr){7-8} \cmidrule(lr){9-10}
& & \textbf{Pass@1} & \textbf{APR} & \textbf{Pass@1} & \textbf{APR} & \textbf{Pass@1} & \textbf{APR} & \textbf{Pass@1} & \textbf{APR} \\
\midrule
\multirow{5}{*}{Open-source models}
& Llama4-Maverick & $26.81$ & $28.72$ & $29.49$ & $32.05$ & $36.36$ & $38.18$ & $36.00$ & $37.14$ \\
& DeepSeek-v3.1 & $32.34$ & $37.23$ & $41.79$ & $48.72$ & $36.73$ & $41.82$ & $34.86$ & $40.00$ \\
& DeepSeek-R1 & $34.89$ & $39.36$ & $39.23$ & $42.31$ & $37.45$ & $43.64$ & $48.57$ & $51.43$ \\
& Kimi-K2 & $23.40$ & $25.53$ & $26.67$ & $29.49$ & $29.09$ & $32.73$ & $29.71$ & $31.43$ \\
& Qwen3-Coder-480B & $30.43$ & $31.91$ & $33.33$ & $35.90$ & $35.27$ & $41.82$ & $39.43$ & $42.86$ \\
\midrule
\multirow{3}{*}{Proprietary models}
& o4-mini & $35.10$ & $37.23$ & $41.56$ & $45.45$ & $41.48$ & $44.44$ & $50.00$ & $58.82$ \\
& GPT-5 & $36.17$ & $39.36$ & $42.31$ & $47.44$ & $\underline{43.64}$ & $45.45$ & $\underline{54.28}$ & $60.00$ \\
& Claude4-Sonnet & $\underline{37.94}$ & $39.36$ & $\underline{49.49}$ & $51.28$ & $42.91$ & $49.09$ & $51.43$ & $54.29$ \\
\midrule
\multirow{5}{*}{RTL-specialized models}
& RTLCoder-v1.1-7B & $0.43$ & $1.06$ & $3.33$ & $5.13$ & $1.09$ & $1.82$ & $0.57$ & $2.86$ \\
& CodeV-7B & $3.83$ & $6.38$ & $5.38$ & $7.69$ & $0.00$ & $0.00$ & $0.00$ & $0.00$ \\
& OriGen-7B & $18.30$ & $21.28$ & $18.97$ & $21.79$ & $14.18$ & $16.36$ & $6.86$ & $11.43$ \\
& CraftRTL-15B & $8.09$ & $11.70$ & $12.31$ & $17.95$ & $13.45$ & $16.36$ & $5.14$ & $8.57$ \\
& ScaleRTL-32B & $25.32$ & $27.66$ & $28.97$ & $33.33$ & $25.82$ & $30.91$ & $30.86$ & $37.14$ \\
\midrule
\multirow{2}{*}{RTL Agents}
& MAGE (Claude4) & --- & $44.68$ & --- & $53.85$ & --- & $50.91$ & --- & $60.00$ \\
& MAGE (\name-Generator) & --- & $\underline{46.81}$ & --- & $\underline{55.13}$ & --- & $\underline{70.91}$ & --- & $\underline{62.86}$ \\
\midrule
\multirow{2}{*}{Ours}
& Generator (standalone) & $\mathbf{39.57}$ & $40.43$ & $\mathbf{49.74}$ & $52.56$ & $\mathbf{65.09}$ & $67.27$ & $\mathbf{56.00}$ & $57.14$ \\
& \name & --- & $\mathbf{80.85}$ & --- & $\mathbf{96.15}$ & --- & $\mathbf{90.91}$ & --- & $\mathbf{91.43}$ \\
\bottomrule
\vspace{-22pt}
\end{tabular}
\end{table*}

\paragraph{\textbf{Baselines}}
We evaluate \name against $14$ SoTA baselines, including frontier open-source models Llama4-Maverick~\cite{meta2024llama4}, DeepSeek-v3.1~\cite{liu2024deepseek}, DeepSeek-R1~\cite{guo2025deepseek}, Kimi-K2~\cite{team2025kimi}, and Qwen3-Coder-480B~\cite{yang2025qwen3}.
Frontier proprietary models comprise 4o-mini~\cite{hurst2024gpt}, GPT-5~\cite{openai2025gpt5}, and Claude4-Sonnet~\cite{anthropic2025claude4}.
For RTL models, we include RTLCoder-v1.1, CodeV~\cite{zhao2025codev}, OriGen~\cite{cui2024origen}, CraftRTL, and ScaleRTL. In addition, as VerilogCoder relies on a custom Verilog parsing tool that is not readily portable to SystemVerilog designs in CVDP, we adopt MAGE as the RTL agentic baseline. To ensure a fair comparison, we use Claude4-Sonnet as the underlying LLM in MAGE, which is the same generic model used in \name.
Our variants are: 
(1) Generator, our RTL-specialized model; 
(2) \name, the full system integrating the specialized Generator with the Reflector and Coordinator.

\paragraph{\textbf{CVDP Benchmark}}
We adopt the CVDP-v1.0.2 benchmark as our primary evaluation suite, which includes four representative RTL design tasks: \emph{code completion} (CVDP-cid002), \emph{specification-to-RTL generation} (CVDP-cid003), \emph{code modification} (CVDP-cid004), and \emph{code debugging} (CVDP-cid016). 
Together, these tasks capture the core challenges faced by RTL designers in practice, covering both generative and corrective design scenarios.
The benchmark contains $94$, $78$, $55$, and $35$ problems under the four categories respectively, covering a diverse range of hardware domains including arithmetic units, on-chip communication protocols, memory hierarchies, DSP kernels, and control logic.
We leave testbench generation and subjective evaluation categories in CVDP to our future work.
Each problem is evaluated across five independent runs.
We report two complementary metrics:
(1) \textbf{Pass@1}, the average success rate across independent runs~\cite{liu2023verilogeval}, and
(2) \textbf{Agentic Pass Rate (APR)}, defined as the ratio of unique solved problems to the total number of problems.
Since agentic methods inherently perform multiple reasoning and correction iterations, comparing them against standalone LLMs solely using Pass@1 can be misleading.
Therefore, we introduce APR to better capture a model’s overall problem-solving coverage across multiple trials, providing a fairer assessment of agentic approaches.
\paragraph{\textbf{Training and Inference}}
\name-Generator is finetuned from Qwen2.5-Coder-32B-Instruct~\cite{hui2024qwen2} using 32 nodes with 8 NVIDIA A100 GPUs each. We employ tensor, pipeline, and context parallelism for efficient large-scale training, and adopt sequence packing over the training data to improve token utilization and speed up the training process. The model is trained for three epochs with a 32,768-token context window and a global batch size of 128 using a cosine-annealing learning rate. The total training process takes approximately 10K GPU-hours, which is a relatively modest cost for training a strong domain-specialized model, especially compared with frontier LLM training, which typically requires multiple orders of magnitude more compute. During inference, \name runs five parallel processes under our parallel scaling strategy, each limited to 30 iterations with a temperature of 1.2. The finetuned Generator is served via \textit{vLLM}~\cite{kwon2023efficient}, while other components use Claude4-Sonnet through its official API.

\begin{figure}[t!]
\begin{center}
	\includegraphics[width=0.95\columnwidth]{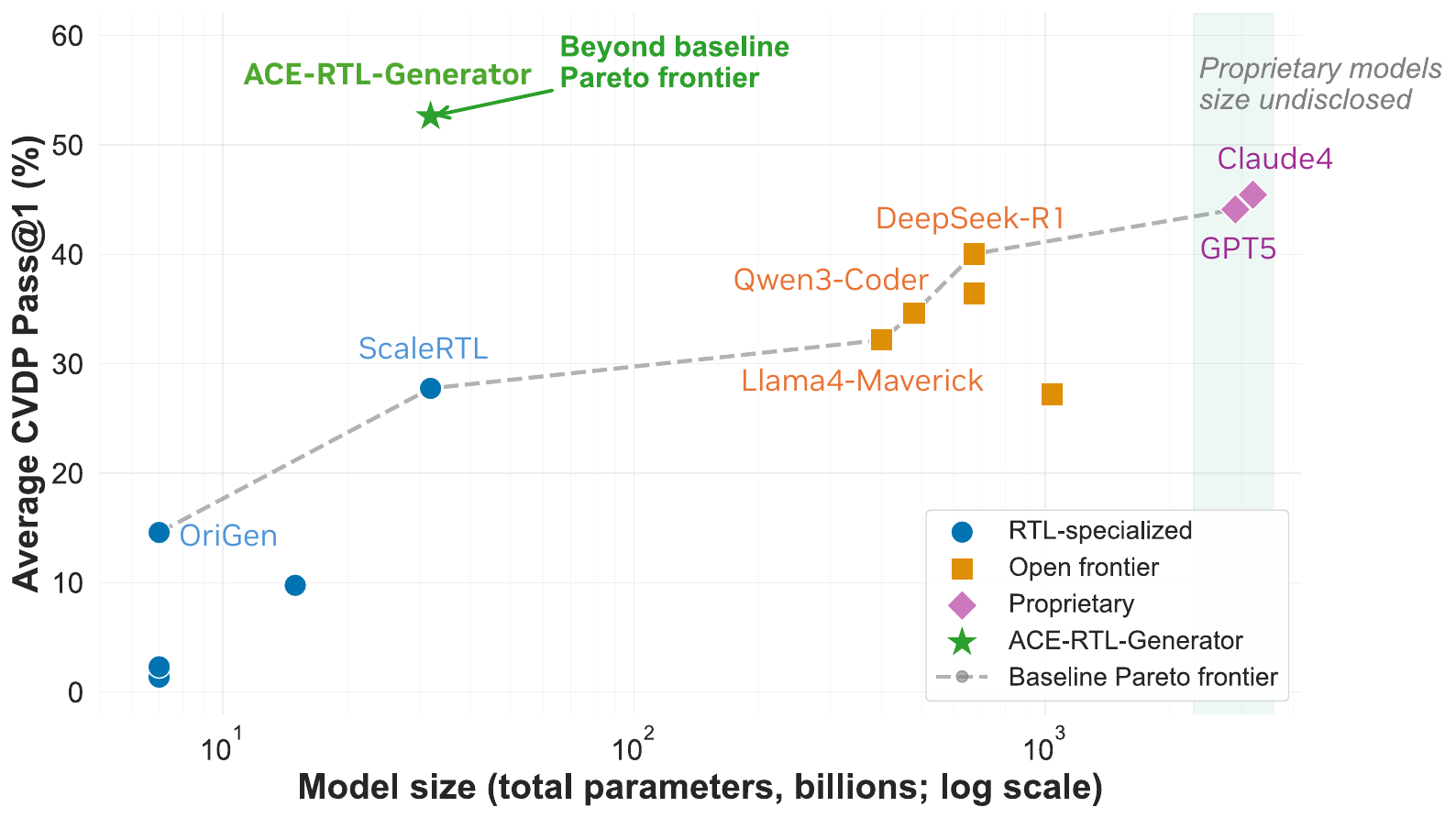}
\vspace{-7pt}
	\caption{Comparison of frontier generic and RTL-specialized LLMs.} 
     \vspace{-22pt}
     \label{figure:comp}
\end{center}
\end{figure}
\subsection{Main Results}
\label{cvdp_results}
Table~\ref{tab:cvdp} shows prior RTL LLMs struggle on CVDP despite performing well on earlier RTL benchmarks, suggesting limited generalization from shorter and simpler RTL data to the more complex designs in CVDP. In contrast, frontier models such as \emph{GPT-5} and \emph{Claude4-Sonnet} show clear improvements and establish strong new baselines, benefiting from broader RTL exposure during large-scale pretraining and stronger reasoning on complex coding tasks.

Our RTL-specialized model, \text{\name}-Generator, further advances the RTL coding frontier, outperforming all standalone model baselines on most tasks. It achieves a Pass@1 of $65.09\%$ on \emph{Code Modification} (cid004), surpassing the SoTA baseline GPT-5 by $21.45\%$. This strong performance stems from our high-quality modification data, created by extracting basic RTL code segments and training the model to augment them with missing features, as discussed in Section~\ref{sec:generator}.

Figure~\ref{figure:comp} compares standalone models in terms of average CVDP Pass@1 across four categories versus model size. In addition to achieving the highest standalone pass rate, \name-Generator also lies beyond the Pareto frontier formed by prior baselines, indicating a particularly favorable trade-off between model scale, which is broadly correlated with training cost, and functional correctness.

For agentic baselines, we evaluate two MAGE variants, using either Claude4 or \text{\name}-Generator as the underlying RTL generation model. In both cases, MAGE achieves only marginal gains over the corresponding standalone model. We attribute this mainly to the absence of a self-evolving context that preserves useful learnings from prior iterations. Without such accumulated debugging history, the agentic loop has limited ability to build on past failures, making it harder to identify root causes and apply targeted fixes over time. A secondary limitation is that MAGE depends on a testbench agent whose outputs can be inaccurate, even when given the golden CVDP test harness, because translating cocotb-based CVDP harnesses into SystemVerilog adds another layer of difficulty on top of the well-known challenge of testbench generation for complex designs~\cite{pinckney2025cvdp}. These results suggest that prior SoTA agentic baselines still struggle to generalize beyond earlier, simpler benchmarks.


Overall, \name achieves the best performance on CVDP, with up to a
$41.02\%$ APR improvement (on cid003) over all baselines. These results show that
combining our RTL-specialized model with iterative context
evolution is highly effective for complex RTL tasks. 

\subsection{Ablation Study}
\label{ablation}
\begin{figure}[t!]
\begin{center}
\vspace{-5pt}
	\includegraphics[width=0.77\columnwidth]{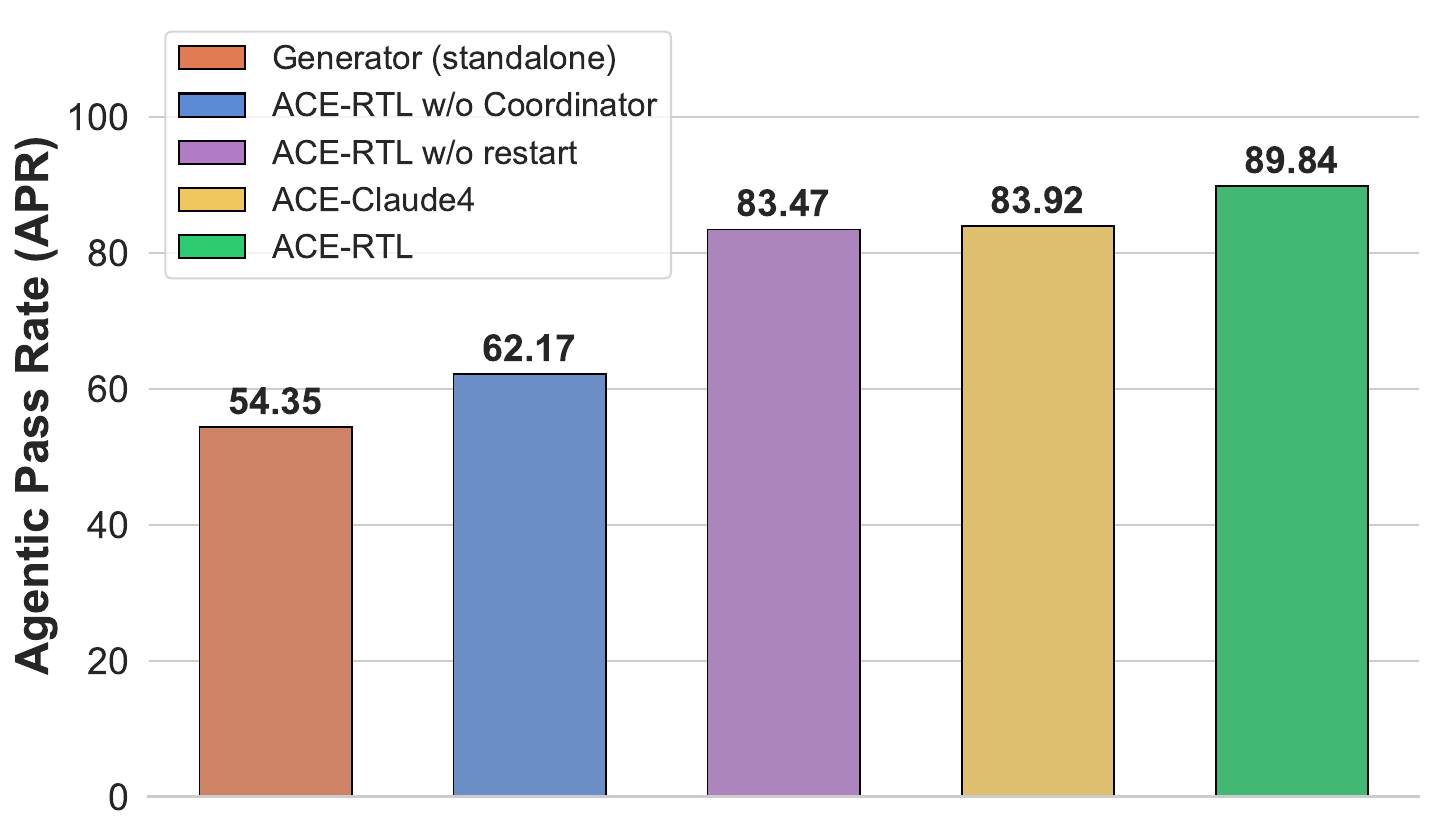}
    \vspace{-9pt}
	\caption{Ablation study on \name components.} 
     \label{figure:ablation}
     \vspace{-23pt}
\end{center}
\end{figure}

Figure~\ref{figure:ablation} reports the averaged APR of several
ablated variants of \name across the four CVDP categories. Relative
to the standalone Generator, \emph{\name\ w/o Coordinator}
already yields a clear improvement, confirming the value of
simulation-guided iterative refinement. The largest gain, however,
comes from enabling the Coordinator in \emph{\name w/o restart},
which maintains an evolving context across iterations using debugging
history from prior attempts. This result shows that preserving
cross-iteration history is a key ingredient in our agentic system,
allowing the Generator to build on prior progress rather than
repeatedly revisiting similar failures.

\begin{figure*}[t]
    \centering

    \begin{minipage}[t]{0.32\textwidth}
        \centering
        \includegraphics[width=\linewidth]{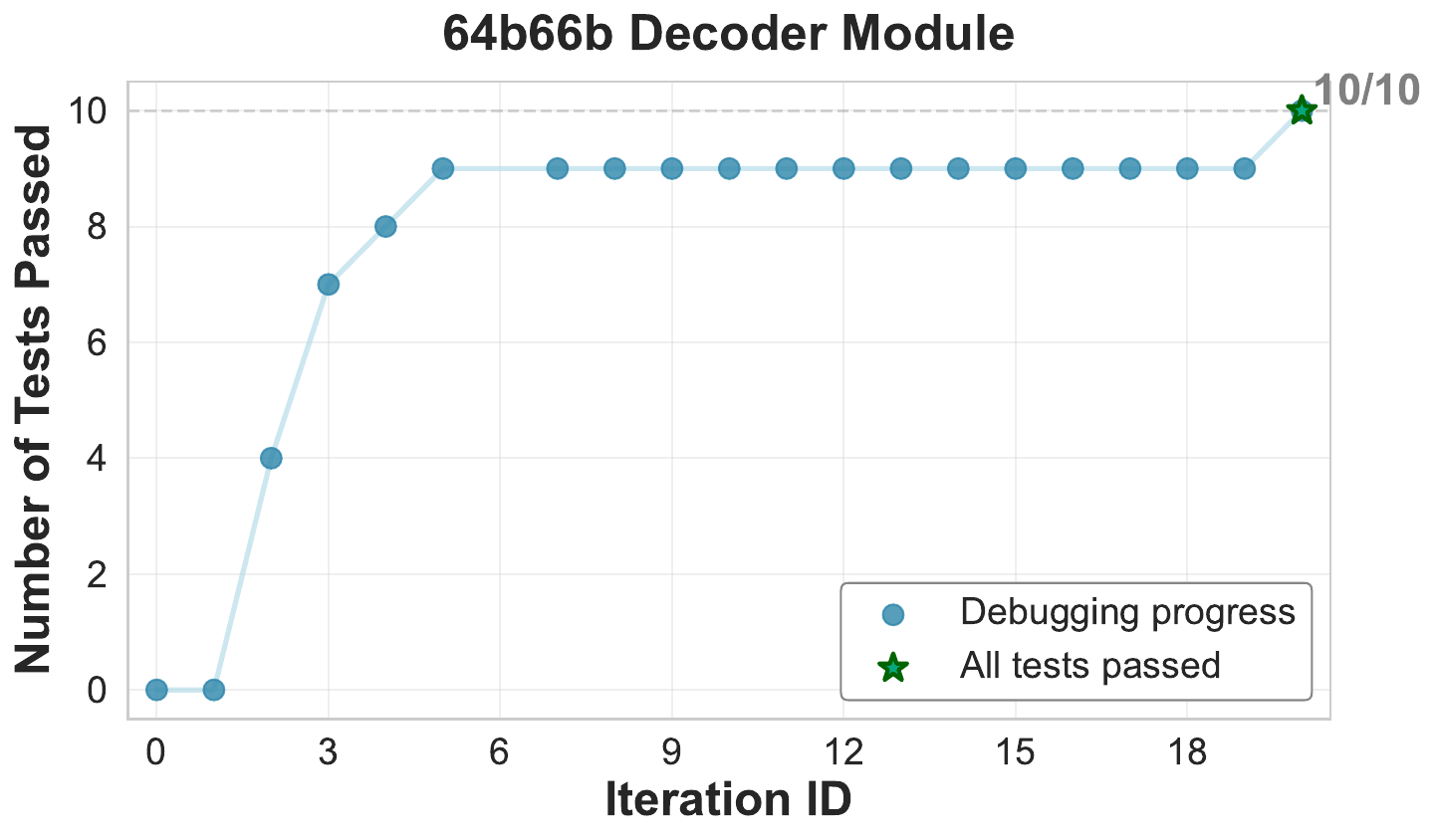}
        \caption{Case Study I.}
        \vspace{-9pt}
        \label{figure:cs2}
    \end{minipage}
    \hfill
    \begin{minipage}[t]{0.32\textwidth}
        \centering
        \includegraphics[width=\linewidth]{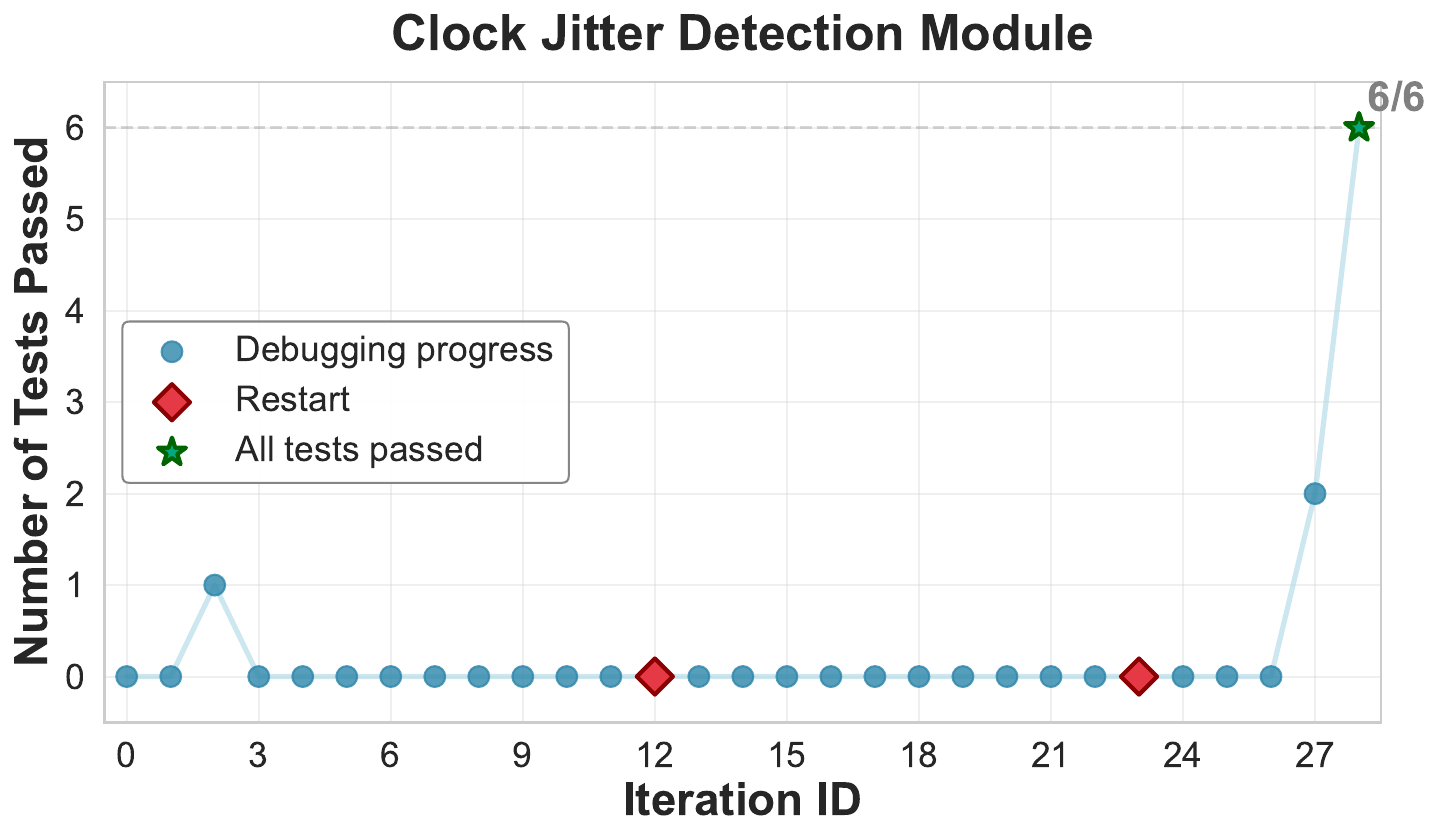}
        \caption{Case Study II.}
        \vspace{-9pt}
        \label{figure:cs3}
    \end{minipage}
    \hfill
    \begin{minipage}[t]{0.32\textwidth}
        \centering
        \includegraphics[width=\linewidth]{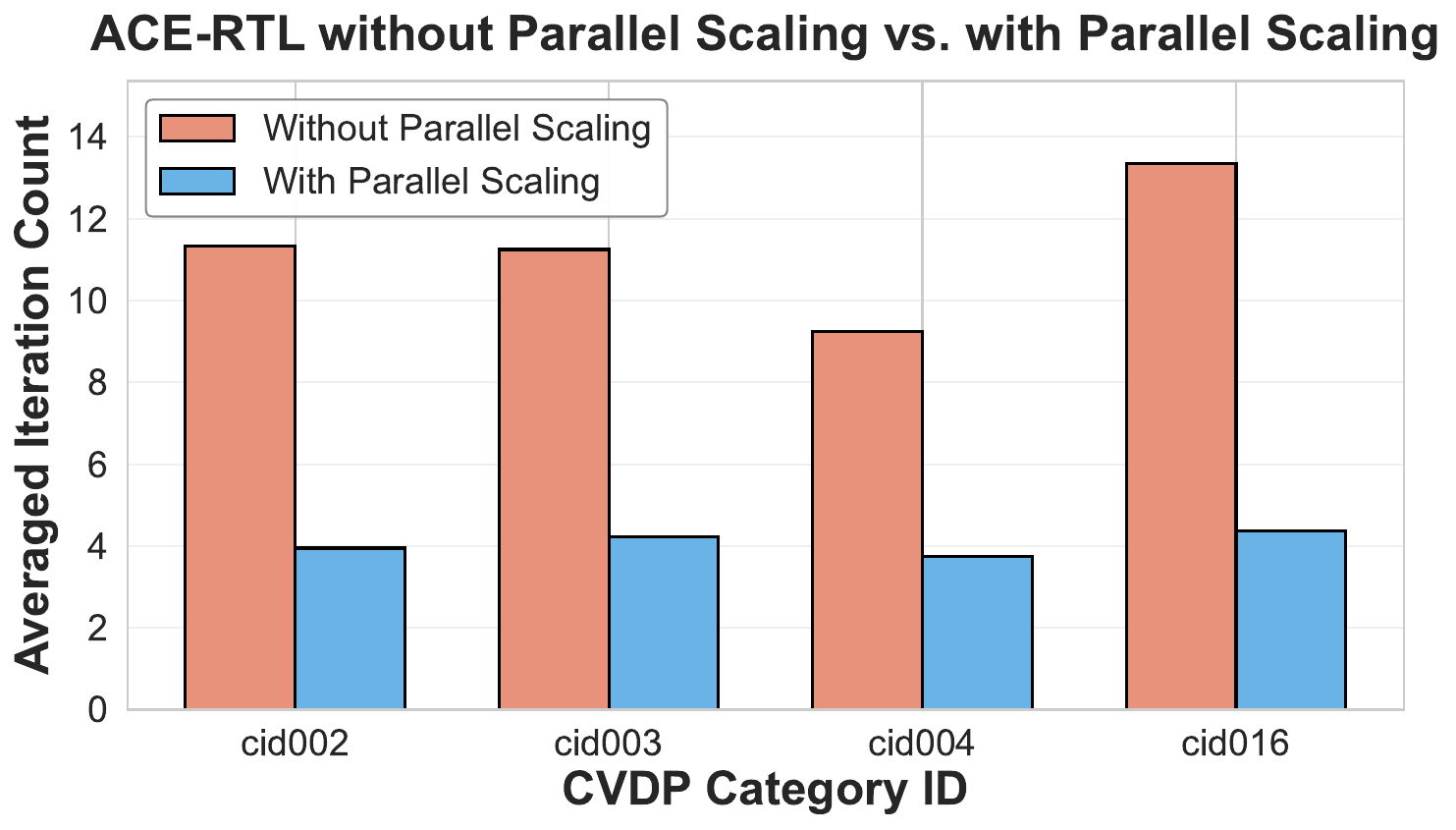}
        \caption{Effect of parallel scaling.}
        \vspace{-9pt}
        \label{figure:speedup}
    \end{minipage}

\end{figure*}

Adding the restart mechanism brings another clear improvement over
\textit{\name w/o restart}, indicating that preserving the evolving
context alone is not always sufficient. Some failed trajectories remain
difficult to recover through incremental local fixes and instead
benefit from regenerating a fresh implementation under the accumulated
debugging context. This behavior is also illustrated in
Figure~\ref{figure:cs3}, where restart helps the system escape
prolonged stagnation.
Finally, ACE-Claude4 uses the same ACE framework but replaces
our RTL Generator with Claude4. Although it
achieves strong performance, the full \name still performs best
overall, demonstrating that agentic context evolution and RTL
specialization are complementary.

\subsection{Case Studies}
\label{cs}
To further demonstrate the effectiveness of \name, we present two representative case studies. The first illustrates how \name moves beyond repeated local fixes to uncover the underlying functional mistake in RTL. The second illustrates how \name identifies unproductive search trajectories and strategically restarts, enabling it to escape local minima and reach a correct RTL implementation.




\textit{\textbf{$\,\,\,\,$ Case Study I ---}} This case study analyzes the 64b/66b Decoder module, which performs bit-aligned extraction of mixed-mode data streams. Figure~\ref{figure:cs2} shows that \name quickly improves functional correctness but remains stuck for many iterations before passing the final test case.
Our Reflector identifies this stagnation and analyzes the discrepancies between expected and actual outputs. By comparing these discrepancies against the provided specification, the Reflector infers that the decoding logic requires an implicit alignment transformation that is not clearly stated in the specification. It then formulates high-level guidance describing this behavioral requirement, which directs the Generator to update the extraction logic accordingly, after which the design immediately passes all remaining tests.


\textit{\textbf{Case Study II ---}} This case study analyzes the Clock Jitter Detection module, which evaluates clock stability by comparing measured edge intervals against a timing threshold. As shown in Figure \ref{figure:cs3}, \name makes early progress but soon enters a long plateau where all attempts fail. The stagnation occurs because each iteration repeats the same flawed behavior: performing jitter detection before a valid baseline interval is established. Although the Generator produces many syntactic variants, they all retain this early-trigger issue and therefore cannot satisfy the required timing semantics.

Our Coordinator recognizes this stalled trajectory by observing that many consecutive iterations exhibit the same assertion pattern with no meaningful improvement. Consequently, the Coordinator triggers a restart and distills high-level insights from the failed attempts, such as the need to validate the first measurement interval before performing detection. It then instructs the Generator to regenerate the design from the specification using these clarified constraints.

The restart yields a simpler implementation that properly validates the measurement interval. After two restarts, the Generator introduces a validity flag that delays jitter detection until the first full interval is observed, restoring correctness. This example illustrates how the Coordinator detects stagnation and uses a \textit{restart mechanism} to guide the system toward a correct RTL implementation.

\begin{table}[t]
\centering
\caption{Impact of PS (parallel scaling) on convergence speed and inference cost, averaged over the four CVDP categories.}
\label{tab:parallel-scaling}
\begin{tabular}{lcccc}
\toprule
Setting & APR $\uparrow$ & Iter. $\downarrow$ & Speedup $\uparrow$ & Rel. Token $\downarrow$ \\
\midrule
w/o PS & 89.84 & 11.30 & 1.00$\times$ & 1.00$\times$ \\
w/ PS  & 89.84 & 4.08  & 2.77$\times$ & 1.12$\times$ \\
\bottomrule
\vspace{-22pt}
\end{tabular}%
\end{table}
\subsection{Efficacy of Parallel Scaling}
\label{para_scale}


Figure~\ref{figure:speedup} shows that parallel scaling consistently
reduces the average number of iterations required to reach a correct
solution across all four CVDP categories, demonstrating that exploring
multiple stochastic trajectories in parallel largely accelerates
convergence. Table~\ref{tab:parallel-scaling} summarizes this trend
over all categories. On average, parallel scaling preserves the same
averaged APR of 89.84 while reducing the iteration count from 11.30 to 4.08,
yielding a 2.77$\times$ speedup. Notably, although parallel scaling
uses five parallel processes, the relative token usage rises to only
1.12$\times$. The reason is that token cost grows nonlinearly across
iterations: in the single-process setting, later iterations carry a much
longer evolving context due to accumulated debugging history, making
each subsequent LLM call substantially more expensive. In contrast,
parallel scaling completes with far fewer iterations per process on
average, so the evolving context in each trajectory remains much
shorter and the average token usage per iteration stays significantly
lower than in the non-scaled setting.
Overall, these results show that the proposed parallel scaling strategy
achieves a more favorable trade-off between convergence speed and
inference cost than the non-scaled baseline.

\subsection{Results on VerilogEval}
\label{verilog_eval}
\begin{table}[t]
\caption{Pass rate results on VerilogEval-Human-v2 --- \name-G denotes the standalone \name-Generator model.}
\label{tab:verilogeval2}
\centering
\resizebox{\columnwidth}{!}{
\begin{tabular}{ccc|ccc}
\toprule
\multicolumn{3}{c|}{\textbf{Pass@1}} & \multicolumn{3}{c}{\textbf{APR}} \\
CraftRTL & Claude4 & \name-G &
VerilogCoder & MAGE & \name \\
\midrule
$68.0$ & $73.0$ & $\mathbf{73.8}$ & $94.2$ & $\mathbf{95.5}$ & $\mathbf{95.5}$ \\
\bottomrule
\vspace{-22pt}
\end{tabular}
}
\end{table}
For completeness, we also evaluate our approach on VerilogEval-Human-v2. We use Pass@1 for standalone models and APR for agentic methods. As shown in Table~\ref{tab:verilogeval2}, \name-G outperforms prior RTL LLMs such as OriGen and CraftRTL. Moreover, \name achieves the highest pass rate among all agentic baselines. These results show that \name is not overfitted to CVDP but instead captures transferable RTL reasoning and coding capabilities, maintaining strong performance on earlier RTL benchmarks. 
\section{Failure Analysis and Future Work}
\label{sec:discussion}
Although \name achieves strong results on CVDP, it does not yet reach
a $100\%$ pass rate. We observe two main sources of remaining failures.
First, some datapoints require fine-grained temporal diagnosis, where
textual simulation feedback alone is insufficient to localize the true
timing-related bug. Second, some failures arise from ambiguity in the
specification, or from subtle mismatches between the intended design
behavior and the behavior emphasized by the test cases. These cases
require not only debugging against observed failures, but also
reasoning about the true design intent described in the specification.

One promising direction for addressing these failures is to incorporate
waveform information, which can reveal temporal mismatches more
directly than textual logs. 
However, we intentionally exclude waveform information in this work to evaluate whether \name can solve RTL tasks primarily through RTL knowledge acquired via domain-specific training and specification-grounded reasoning, using only limited non-waveform simulation feedback.
Although waveform-driven debugging often boosts the pass rate of LLM agents, it also risks overfitting to observed test behavior while underutilizing the specification itself, especially in cases with specification ambiguity or specification-test mismatch. We therefore view waveform information not as a substitute for specification-grounded reasoning, but as a potentially useful auxiliary signal that should be integrated into \name in a careful and principled way to avoid overfitting to test cases. We leave such integration as an important future direction.


\section{Conclusion}
This work introduces \name, a framework that integrates an RTL model with ACE and achieves SoTA results on multiple RTL benchmarks. An interesting future direction is to automate testbench generation and explore waveform-enhanced agentic reasoning.


\bibliographystyle{plain}
\bibliography{ref}

@misc{liu2023chipnemo,
  title  = {ChipNemo: Domain-Adapted LLMs for Chip Design},
  author = {Liu, Mingjie and Ene, Teodor-Dumitru and Kirby, Robert and Cheng, Chris and Pinckney, Nathaniel and Liang, Rongjian and Alben, Jonah and Anand, Himyanshu and Banerjee, Sanmitra and Bayraktaroglu, Ismet and others},
  year   = {2023},
  eprint = {2311.00176},
  archivePrefix = {arXiv}
}

@inproceedings{deng2025chipalign,
  title={Chipalign: Instruction alignment in large language models for chip design via geodesic interpolation},
  author={Deng, Chenhui and Bai, Yunsheng and Ren, Haoxing},
  booktitle={62nd Design Automation Conference (DAC)},
  publisher = {IEEE},
  address   = {San Francisco, CA, USA},
  pages={1--7},
  year={2025}
}

@ARTICLE{liu2024rtlcoderb,
  author={Liu, Shang and Fang, Wenji and Lu, Yao and Wang, Jing and Zhang, Qijun and Zhang, Hongce and Xie, Zhiyao},
  journal={IEEE Transactions on Computer-Aided Design of Integrated Circuits and Systems}, 
  title={RTLCoder: Fully Open-Source and Efficient LLM-Assisted RTL Code Generation Technique}, 
  year={2025},
  volume={44},
  number={4},
  pages={1448-1461},
  doi={10.1109/TCAD.2024.3483089}}

@INPROCEEDINGS{liu2024rtlcodera,
  author={Liu, Shang and Fang, Wenji and Lu, Yao and Zhang, Qijun and Zhang, Hongce and Xie, Zhiyao},
  booktitle={2024 IEEE LLM Aided Design Workshop (LAD)}, 
  title={RTLCoder: Outperforming GPT-3.5 in Design RTL Generation with Our Open-Source Dataset and Lightweight Solution}, 
  year={2024},
  volume={},
  number={},
  pages={1-5},
  doi={10.1109/LAD62341.2024.10691788},
  publisher={IEEE},
  address={2024},
}

@misc{liu2024craftrtlhigh,
      title={CraftRTL: High-quality Synthetic Data Generation for Verilog Code Models with Correct-by-Construction Non-Textual Representations and Targeted Code Repair}, 
      author={Mingjie Liu and Yun-Da Tsai and Wenfei Zhou and Haoxing Ren},
      year={2024},
      eprint={2409.12993},
      archivePrefix={arXiv},
      primaryClass={cs.AR},
      url={https://arxiv.org/abs/2409.12993}, 
}

@misc{deng2025scalertl,
      title={ScaleRTL: Scaling LLMs with Reasoning Data and Test-Time Compute for Accurate RTL Code Generation}, 
      author={Chenhui Deng and Yun-Da Tsai and Guan-Ting Liu and Zhongzhi Yu and Haoxing Ren},
      year={2025},
      booktitle = {2025 ACM/IEEE International Symposium on Machine Learning for CAD (LMCAD)},      
      eprint={2506.05566},
      archivePrefix={arXiv},
      primaryClass={cs.AR},
      url={https://arxiv.org/abs/2506.05566}, 
}

@misc{pinckney2024verilogeval,
      title={Revisiting VerilogEval: Newer LLMs, In-Context Learning, and Specification-to-RTL Tasks}, 
      author={Nathaniel Pinckney and Christopher Batten and Mingjie Liu and Haoxing Ren and Brucek Khailany},
      year={2024},
      eprint={2408.11053},
      archivePrefix={arXiv},
      primaryClass={cs.SE},
      url={https://arxiv.org/abs/2408.11053}, 
}

@inproceedings{lu2024rtllm,
  author={Lu, Yao and Liu, Shang and Zhang, Qijun and Xie, Zhiyao},
  booktitle={2024 29th Asia and South Pacific Design Automation Conference (ASP-DAC)}, 
  title={RTLLM: An Open-Source Benchmark for Design RTL Generation with Large Language Model}, 
  year={2024},
  pages={722-727},
  publisher={IEEE},
  }

@inproceedings{Ho2025verilogcoder,
author = {Ho, Chia-Tung and Ren, Haoxing and Khailany, Brucek},
title = {VerilogCoder: autonomous verilog coding agents with graph-based planning and abstract syntax tree (AST)-based waveform tracing tool},
year = {2025},
isbn = {978-1-57735-897-8},
publisher = {AAAI Press},
url = {https://doi.org/10.1609/aaai.v39i1.32007},
doi = {10.1609/aaai.v39i1.32007},
booktitle = {Proceedings of the Thirty-Ninth AAAI Conference on Artificial Intelligence and Thirty-Seventh Conference on Innovative Applications of Artificial Intelligence and Fifteenth Symposium on Educational Advances in Artificial Intelligence},
articleno = {34},
numpages = {8},
series = {AAAI'25/IAAI'25/EAAI'25},
address = {AAAI},
}

@misc{zhao2024mage,
      title={MAGE: A Multi-Agent Engine for Automated RTL Code Generation}, 
      author={Yujie Zhao and Hejia Zhang and Hanxian Huang and Zhongming Yu and Jishen Zhao},
      year={2024},
      eprint={2412.07822},
      archivePrefix={arXiv},
      primaryClass={cs.AR},
      url={https://arxiv.org/abs/2412.07822}, 
}

@misc{allam2024rtlrepo,
      title={RTL-Repo: A Benchmark for Evaluating LLMs on Large-Scale RTL Design Projects}, 
      author={Ahmed Allam and Mohamed Shalan},
      year={2024},
      eprint={2405.17378},
      archivePrefix={arXiv},
      primaryClass={cs.LG}
}

@misc{pinckney2025cvdp,
      title={Comprehensive Verilog Design Problems: A Next-Generation Benchmark Dataset for Evaluating Large Language Models and Agents on RTL Design and Verification}, 
      author={Nathaniel Pinckney and Chenhui Deng and Chia-Tung Ho and Yun-Da Tsai and Mingjie Liu and Wenfei Zhou and Brucek Khailany and Haoxing Ren},
      year={2025},
      eprint={2506.14074},
      archivePrefix={arXiv},
      primaryClass={cs.LG},
      url={https://arxiv.org/abs/2506.14074}, 
}

@article{anthropic2024claude,
  title={The claude 3 model family: Opus, sonnet, haiku},
  author={Anthropic, AI},
  journal={Claude-3 Model Card},
  volume={1},
  number={1},
  pages={4},
  year={2024}
}

@techreport{anthropic2025claude4,
  title        = {Claude 4 System Card: Claude Opus 4 \& Claude Sonnet 4},
  author       = {Anthropic},
  year         = {2025},
  month        = {May},
  institution  = {Anthropic},
  url          = {https://www-cdn.anthropic.com/6be99a52cb68eb70eb9572b4cafad13df32ed995.pdf}
}

@techreport{openai2025gpt5,
  title        = {GPT-5 System Card},
  author       = {OpenAI},
  year         = {2025},
  month        = {August},
  institution  = {OpenAI},
  url          = {https://cdn.openai.com/gpt-5-system-card.pdf}
}

@misc{zhang2025aspen,
  title={ASPEN: LLM-Guided E-Graph Rewriting for RTL Datapath Optimization},
  author={Zhang, Niansong and Deng, Chenhui and Kuehn, Johannes Maximilian and Ho, Chia-Tung and Yu, Cunxi and Zhang, Zhiru and Ren, Haoxing},
  booktitle={2025 ACM/IEEE 7th Symposium on Machine Learning for CAD (MLCAD)},
  year={2025},
  organization={IEEE}
}

@misc{liu2023verilogeval,
  title={{VerilogEval:} Evaluating Large Language Models for Verilog Code Generation},
  author={Liu, Mingjie and Pinckney, Nathaniel and Khailany, Brucek and Ren, Haoxing},
  booktitle={2023 IEEE/ACM International Conference on Computer-Aided Design (ICCAD)}, 
  year={2023}
}

@misc{guo2025deepseek,
  title={Deepseek-r1: Incentivizing reasoning capability in llms via reinforcement learning},
  author={Guo, Daya and Yang, Dejian and Zhang, Haowei and Song, Junxiao and Zhang, Ruoyu and Xu, Runxin and Zhu, Qihao and Ma, Shirong and Wang, Peiyi and Bi, Xiao and others},
  journal={arXiv preprint arXiv:2501.12948},
  year={2025}
}

@article{williams2002icarus,
  title={Icarus verilog: open-source verilog more than a year later},
  author={Williams, Stephen and Baxter, Michael},
  journal={Linux Journal},
  volume={2002},
  number={99},
  pages={3},
  year={2002},
  publisher={Belltown Media Houston, TX}
}

@article{chung2024scaling,
  title={Scaling instruction-finetuned language models},
  author={Chung, Hyung Won and Hou, Le and Longpre, Shayne and Zoph, Barret and Tay, Yi and Fedus, William and Li, Yunxuan and Wang, Xuezhi and Dehghani, Mostafa and Brahma, Siddhartha and others},
  journal={Journal of Machine Learning Research},
  volume={25},
  number={70},
  pages={1--53},
  year={2024}
}

@misc{meta2024llama4,
  author       = {Meta AI},
  title        = {LLaMA 4: Multimodal Intelligence},
  year         = {2024},
  url          = {https://ai.meta.com/blog/llama-4-multimodal-intelligence/},
  note         = {Accessed: 2025-11-16}
}

@misc{liu2024deepseek,
  title={Deepseek-v3 technical report},
  author={Liu, Aixin and Feng, Bei and Xue, Bing and Wang, Bingxuan and Wu, Bochao and Lu, Chengda and Zhao, Chenggang and Deng, Chengqi and Zhang, Chenyu and Ruan, Chong and others},
  journal={arXiv preprint arXiv:2412.19437},
  year={2024}
}

@misc{team2025kimi,
  title={Kimi k2: Open agentic intelligence},
  author={Team, Kimi and Bai, Yifan and Bao, Yiping and Chen, Guanduo and Chen, Jiahao and Chen, Ningxin and Chen, Ruijue and Chen, Yanru and Chen, Yuankun and Chen, Yutian and others},
  journal={arXiv preprint arXiv:2507.20534},
  year={2025}
}

@misc{yang2025qwen3,
  title={Qwen3 technical report},
  author={Yang, An and Li, Anfeng and Yang, Baosong and Zhang, Beichen and Hui, Binyuan and Zheng, Bo and Yu, Bowen and Gao, Chang and Huang, Chengen and Lv, Chenxu and others},
  journal={arXiv preprint arXiv:2505.09388},
  year={2025}
}

@misc{hurst2024gpt,
  title={Gpt-4o system card},
  author={Hurst, Aaron and Lerer, Adam and Goucher, Adam P and Perelman, Adam and Ramesh, Aditya and Clark, Aidan and Ostrow, AJ and Welihinda, Akila and Hayes, Alan and Radford, Alec and others},
  journal={arXiv preprint arXiv:2410.21276},
  year={2024}
}

@misc{hui2024qwen2,
  title={Qwen2. 5-coder technical report},
  author={Hui, Binyuan and Yang, Jian and Cui, Zeyu and Yang, Jiaxi and Liu, Dayiheng and Zhang, Lei and Liu, Tianyu and Zhang, Jiajun and Yu, Bowen and Lu, Keming and others},
  journal={arXiv preprint arXiv:2409.12186},
  year={2024}
}

@misc{kwon2023efficient,
  title={Efficient memory management for large language model serving with pagedattention},
  author={Kwon, Woosuk and Li, Zhuohan and Zhuang, Siyuan and Sheng, Ying and Zheng, Lianmin and Yu, Cody Hao and Gonzalez, Joseph and Zhang, Hao and Stoica, Ion},
  booktitle={Proceedings of the 29th symposium on operating systems principles},
  pages={611--626},
  year={2023}
}

@misc{jimenez2023swe,
  title={Swe-bench: Can language models resolve real-world github issues?},
  author={Jimenez, Carlos E and Yang, John and Wettig, Alexander and Yao, Shunyu and Pei, Kexin and Press, Ofir and Narasimhan, Karthik},
  journal={arXiv preprint arXiv:2310.06770},
  year={2023}
}

@misc{cursor2025,
  author       = {{Cursor AI}},
  title        = {{Cursor: The Best Way to Code with AI}},
  howpublished = {\url{https://cursor.com/}},
  year         = {2025},
  note         = {Accessed: 2025-02-10}
}

@misc{claudecode2025,
  author       = {{Anthropic}},
  title        = {{Claude Code: Best Practices for Agentic Coding}},
  howpublished = {\url{https://www.anthropic.com/engineering/claude-code-best-practices}},
  year         = {2025},
  note         = {Accessed: 2025-02-10}
}

@misc{yao2024rtlrewriter,
  title={Rtlrewriter: Methodologies for large models aided rtl code optimization},
  author={Yao, Xufeng and Wang, Yiwen and Li, Xing and Lian, Yingzhao and Chen, Ran and Chen, Lei and Yuan, Mingxuan and Xu, Hong and Yu, Bei},
  booktitle={Proceedings of the 43rd IEEE/ACM International Conference on Computer-Aided Design},
  year={2024}
}

@misc{zhao2025codev,
  title={Codev: Empowering llms with hdl generation through multi-level summarization},
  author={Zhao, Yang and Huang, Di and Li, Chongxiao and Jin, Pengwei and Song, Muxin and Xu, Yinan and Nan, Ziyuan and Gao, Mingju and Ma, Tianyun and Qi, Lei and others},
  journal={IEEE Transactions on Computer-Aided Design of Integrated Circuits and Systems},
  year={2025},
  publisher={IEEE}
}

@misc{cui2024origen,
  title={Origen: Enhancing rtl code generation with code-to-code augmentation and self-reflection},
  author={Cui, Fan and Yin, Chenyang and Zhou, Kexing and Xiao, Youwei and Sun, Guangyu and Xu, Qiang and Guo, Qipeng and Liang, Yun and Zhang, Xingcheng and Song, Demin and others},
  booktitle={Proceedings of the 43rd IEEE/ACM International Conference on Computer-Aided Design},
  year={2024}
}

@misc{wang2025symrtlo,
  title={SymRTLO: Enhancing RTL Code Optimization with LLMs and Neuron-Inspired Symbolic Reasoning},
  author={Wang, Yiting and Ye, Wanghao and Guo, Ping and He, Yexiao and Wang, Ziyao and Tian, Bowei and He, Shwai and Sun, Guoheng and Shen, Zheyu and Chen, Sihan and others},
  journal={Conference on Neural Information Processing Systems (NeurIPS)},
  year={2025}
}

@misc{openai2024gpt4turbo,
  author       = {OpenAI},
  title        = {New models and developer products announced at DevDay},
  year         = {2024},
  howpublished = {\url{https://openai.com/index/new-models-and-developer-products-announced-at-devday/}},
  note         = {Accessed: 2025-11-19}
}

@article{gu2024survey,
  title={A survey on llm-as-a-judge},
  author={Gu, Jiawei and Jiang, Xuhui and Shi, Zhichao and Tan, Hexiang and Zhai, Xuehao and Xu, Chengjin and Li, Wei and Shen, Yinghan and Ma, Shengjie and Liu, Honghao and others},
  journal={The Innovation},
  year={2024},
  publisher={Elsevier}
}

@article{yang2023rethinking,
  title={Rethinking benchmark and contamination for language models with rephrased samples},
  author={Yang, Shuo and Chiang, Wei-Lin and Zheng, Lianmin and Gonzalez, Joseph E and Stoica, Ion},
  journal={arXiv preprint arXiv:2311.04850},
  year={2023}
}

\end{document}